\documentclass[journal]{IEEEtran}
\ifCLASSINFOpdf
\else
\fi
\usepackage[cmex10]{amsmath}
\usepackage{upgreek}
\usepackage{booktabs}
\usepackage{epsfig}
\usepackage{latexsym}
\usepackage{multirow}
\usepackage{stfloats}
\usepackage{epstopdf}
\usepackage{color}  
\usepackage{tabularx} 
\usepackage{amssymb}
\usepackage{enumerate}
\graphicspath{{./Figures/}}
\usepackage{color}
\usepackage{bbm}
\usepackage{bm}
\usepackage{cite}
\usepackage[tight,footnotesize]{subfigure}
\usepackage{balance}
\usepackage{mathrsfs}
\usepackage{verbatim}
\usepackage{dsfont}
\usepackage{verbatim}
\usepackage{tikz}
\usepackage{setspace}
\usepackage{diagbox}
\usepackage{caption}
\usepackage[framemethod=tikz]{mdframed}
\usepackage{multicol}
\usepackage{environ}
\usepackage{tikz}
\usepackage{amsmath}
\usepackage{stfloats}

\begin{document}
\title{Hybrid Beamforming for Terahertz Wireless Communications: Challenges, Architectures, and Open Problems}
\author{Chong Han,~\IEEEmembership{Member,~IEEE}, Longfei Yan, and Jinhong Yuan,~\IEEEmembership{Fellow,~IEEE}
\thanks{
Chong Han and Longfei Yan are with the Terahertz Wireless Communications (TWC) laboratory, University of Michigan-Shanghai Jiao Tong University Joint Institute, Shanghai Jiao Tong University, Shanghai 200240, China (e-mail: chong.han@sjtu.edu.cn; longfei.yan@sjtu.edu.cn).

Jinhong Yuan is with the School of Electrical Engineering and Telecommunications, University of New South Wales, Sydney, NSW 2052, Australia (e-mail: j.yuan@unsw.edu.au).}}
\maketitle
	
\begin{abstract}
Terahertz (THz) communications are regarded as a pillar technology for the sixth generation (6G) wireless systems, by offering multi-ten-GHz bandwidth. 
To overcome the short transmission distance and huge propagation loss, ultra-massive (UM) MIMO systems that employ sub-millimeter wavelength antennas array are proposed to enable an enticingly high array gain. In the UM-MIMO systems, hybrid beamforming stands out for its great potential in promisingly high data rate and reduced power consumption.
In this paper, challenges and features of the THz hybrid beamforming design are investigated, in light of the distinctive THz peculiarities. 
Specifically, we demonstrate that the spatial degree-of-freedom (SDoF) is less than 5, which is caused by the extreme sparsity of the THz channel. The blockage problem caused by the huge reflection and scattering losses, as high as 15 dB or over, is studied. 
Moreover, we analyze the challenges led by the array containing 1024 or more antennas, including the requirement for intelligent subarray architecture, strict energy efficiency, and propagation characterization based on spherical-wave propagation mechanisms.
Owning up to hundreds of GHz bandwidth, beam squint effect could cause over 5~dB array gain loss, when the fractional bandwidth exceeds 10\%.
Inspired by these facts, three novel THz-specific hybrid beamforming architectures are presented, including widely-spaced multi-subarray, dynamic array-of-subarrays, and true-time-delay-based 
architectures. 
We also demonstrate the potential data rate, power consumption, and array gain capabilities for THz communications. As a roadmap of THz hybrid beamforming design, multiple open problems and potential research directions are elaborated.
\end{abstract}
\IEEEpeerreviewmaketitle
\section{Introduction}
\IEEEPARstart{W}{ith} the commercialization and global deployment of the fifth generation (5G) wireless communications, increasing research attention and efforts are migrating to development of sixth generation (6G) wireless communications. 
As an important key performance indicator for 6G, the peak data rate is expected to reach 1~Tbps~\cite{AKYILDIZ201646}. 
Although the trend of exploring higher frequency spectrum is clear, it is still difficult for
millimeter wave (mmWave) systems to support Tbps data rate, due to the limited consecutive available bandwidth up to several GHz. 
Lying between the mmWave and infrared spectrum, the Terahertz (THz) band spans over 0.1-10~THz, which owns several tens of GHz consecutive available bandwidth and is envisioned as an enabling technology for 6G wireless systems~\cite{sarieddeen2019terahertzband}.

However, the THz band suffers from huge propagation loss due to the small effective area of the THz antenna, which is proportional to the square of the wavelength~\cite{6998944}. The huge propagation loss significantly limits the wireless communication distance.
Fortunately, due to the sub-millimeter wavelength, ultra-massive antennas can be arranged, e.g., 1024 antennas, to realize THz ultra-massive multi-input multi-output (UM-MIMO) systems~\cite{AKYILDIZ201646,7402270}, which can steer a narrowbeam and generate a high array gain to address the distance constraint. 

In terms of beamforming architectures, the fully-digital and analog are two conventional selections.
On one hand, in the fully-digital architecture, each antenna owns a dedicated RF chain and DAC/ADC, which are power-hungry~\cite{Yuan2020TCOM,1}. Since the number of antennas can be prohibitively high in the THz UM-MIMO systems, hardware complexity and power consumption of the fully-digital beamforming architecture are unbearable for practical use~\cite{8733134,111}. 
On the other hand, only one RF chain and DAC/ADC are used to control all antennas through phase shifters in the analog beamforming architecture. 
While the power consumption and hardware complexity are substantially reduced, the analog beamforming architecture can only support one data stream, which strictly limits the data rate and the number of users. 
As a combination of these two architectures, hybrid beamforming is proposed. With a limited number of RF chains and DAC/ADCs, the hybrid beamforming can achieve comparable data rate with fully-digital architecture, while with reduced hardware complexity and power consumption. 
Consequently, the hybrid beamforming has been treated as an appealing technology in THz UM-MIMO systems~\cite{Yuan2020TCOM}. Although the hybrid beamforming technologies have been investigated extensively for the microwave and mmWave frequencies~\cite{1,7389996,7445130,7397861}, the peculiarities of the THz UM-MIMO systems bring many new challenges for the design of THz hybrid beamforming~\cite{Yuan2020TCOM,8733134,7959180}.

In this paper, we first analyze the distinctive features and challenges of the THz hybrid beamforming. Specifically, we show that the spatial degree-of-freedom (SDoF) can be less than 5, which is caused by the extreme sparsity of the THz channel as a result of blockage and huge reflection and scattering losses, e.g., larger than 15 dB. 
Moreover, we analyze the challenges led by the array containing as many as 1024 or more antennas, including the demand for intelligent subarray architecture, strict energy efficiency, and propagation characterization based on spherical-wave propagation mechanisms.
We illustrate that owning up to hundreds of GHz bandwidth, beam squint effect could cause over 5~dB array gain loss, when the fractional bandwidth exceeds 10\%. Then, we survey two traditional hybrid beamforming architectures and reveal their limitations. 
In light of the challenges and being motivated by the constraints of existing approaches, we investigate three interesting THz-specific hybrid beamforming architectures. Furthermore, typical results are presented to illustrate the data rate, power consumption, and array gain performance of these hybrid beamforming designs in the THz band. Last but not least, multiple open problems and research directions are discussed for 6G THz hybrid beamforming design.

\section{Challenges of THz Hybrid Beamforming}
\label{section_challenges}
In this section, the challenges and unique features of the THz hybrid beamforming from the channel and UM-MIMO systems perspectives are elaborated.

\subsection{Challenges from THz Channel Perspectives}
\subsubsection{\textbf{Channel sparsity and low SDoF}}
Owing to the sub-millimeter wavelength, the THz band suffers huge scattering and diffraction losses, compared to the microwave and mmWave frequencies. 
Therefore, the THz channel is usually composed by a line-of-sight (LoS) path and a few reflection paths~\cite{6998944}. The number of multipath is very limited, e.g., typically less than 5, and the THz channel is much sparser than the microwave and mmWave channels~\cite{Yuan2020TCOM}. Since the SDoF is upper-bounded by the number of multipath, the small SDoF restricts the spatial multiplexing gain of the THz UM-MIMO systems. Even with a high array gain generated by ultra-massive antennas, the poor multiplexing gain still significantly limits the potential of the THz UM-MIMO systems, particularly the data rate.

\subsubsection{\textbf{Blockage problem}}
As aforementioned, the THz channel is composed by a LoS path and several reflection paths. Due to the huge reflection loss, the LoS path is significantly stronger than the reflection paths, e.g., more than 15 dB~\cite{6998944}.
The resulting LoS dominance makes the THz wireless links very vulnerable to blockage. Compared to the mmWave frequencies, on one hand, the reflection paths of the THz channel are much weaker. On the other hand, the data rate requirement of the THz systems is usually much higher than the mmWave systems. In light of these, when the LoS path is blocked, the remaining reflection paths might be too weak to support high data rates and the THz link faces a more severe blockage problem than the mmWave link.

\subsection{Challenges from THz UM-MIMO Systems Perspectives}
\subsubsection{\textbf{Large-scale antenna array}}
Since the use of huge number of antennas in THz UM-MIMO systems, i.e., more than 1024 antennas, the power consumption becomes a non-negligible problem.
On one hand, due to the high frequency of the THz band, the power consumption of the individual hardware device is ultra-high, e.g., power amplifier and phase shifter consume 60 mW and 42 mW, respectively~\cite{111}. On the other hand, ultra-massive antennas in the THz UM-MIMO systems result in the use of a large quantity of devices. As a result of these two reasons, the THz UM-MIMO systems suffer prohibitively high power consumption, e.g., several tens of watts~\cite{111}. Hence, the reduction of power consumption is critical to the THz hybrid beamforming design, while achieving high data rates.

\subsubsection{\textbf{Beam squint effect}}
Note that the fractional bandwidth is defined as the ratio between the bandwidth and the central frequency. In traditional microwave frequencies, due to the scarcity of the spectrum resource, the fractional bandwidth is usually very small, e.g., 20 MHz/2.4~GHz=0.83\% at 2.4~GHz. While for mmWave and THz bands, especially the THz band, the fractional bandwidth is much larger, e.g., 50~GHz/300~GHz=16.7\% at 300~GHz, which causes a severe beam squint problem in THz hybrid beamforming architectures.
During the beamforming process, the signals on antennas need phase shifts to be concentrated, which are related to the carrier frequency. However, most of the existing hybrid beamforming architectures are based on the phase shifter, which is a frequency-independent device and tunes the same phase shift for signals with different frequencies. The phase shifter can only tune the ``correct'' phase shift for one frequency point, while for other frequencies, there is a phase error of the signals~\cite{7959180}. As a result, the generated beams are squint and the array gains are reduced, which is the so-called beam squint problem.
In traditional narrowband hybrid beamforming architecture with limited fractional bandwidth, the performance degradation caused by the beam squint problem is acceptable. However, for THz hybrid beamforming architectures with large fractional bandwidth, the beam squint problem can reduce the array gain significantly, e.g., by 5 dB, and needs to be treated seriously~\cite{7959180}.

\subsubsection{\textbf{Spherical-wave propagation}}
Most existing studies about MIMO systems assume the planar-wave propagation, which is an approximation and simplification of the spherical-wave propagation. The planar-wave assumption is accurate when the communication distance is longer than the Rayleigh distance of the antenna array, i.e., the border between the radiating near-field and far-field of the array~\cite{8356240}. The Rayleigh distance equals to the square of the array size divided by half of wavelength. For microwave and mmWave systems, this assumption is reasonable since the Rayleigh distance is pretty short. For instance, considering the array size as 0.1m, the Rayleigh distance is only 0.4m and 4m when the frequencies are 6 GHz and 60 GHz. However, when the frequency comes to the THz band, e.g., 1 THz, the Rayleigh distance grows to 67m.
Hence, the communication distance might be smaller than the Rayleigh distance in the THz band and the planar-wave assumption is invalid. Instead, in the THz UM-MIMO channels, the spherical-wave propagation needs to be considered, which brings additional difficulties in channel modeling and estimation design and further influences the hybrid beamforming.

\section{Traditional Hybrid Beamforming Architectures and Limitations}
\label{section_general}
We first introduce the traditional hybrid beamforming architectures and algorithms in the literature~\cite{1,7445130,7389996,7397861}. Then, by considering the THz-specific challenges, we analyze the limitations of these architectures if being implemented in THz UM-MIMO systems. 

\subsection{Existing Hybrid Beamforming Architectures and Algorithms}
Among hybrid beamforming architectures, the fully-connected (FC) architecture has been investigated extensively~\cite{1}.
In the FC architecture, each RF chain controls all antennas through phase shifters. The number of phase shifters equals to the product of the number of antennas and the number of RF chains.
However, owing to the use of extensive phase shifters, hardware complexity as well as the power consumption of FC architecture are still unacceptably high for practical implementation. 
To tackle this issue, the array-of-subarrays (AoSA) architecture is desirable~\cite{7445130}, in which each RF chain connects to only a subset of antennas, i.e., a subarray, through phase shifters. The resulting number of phase shifters equals to the number of antennas. To this end, hardware complexity and power consumption of the AoSA architecture are substantially lower than the FC architecture. 

\textit{\textbf{Hybrid beamforming algorithm design}}: Most of the existing design target of hybrid beamforming architectures is to maximize the data rate, which is realized by solving the analog and digital beamforming matrices.
For FC architecture, a two-step hybrid beamforming algorithm~\cite{7389996} can be exploited to maximize the data rate, by decoupling the optimization of the analog and digital beamforming matrices. For AoSA architecture, a successive-interference-cancellation (SIC) based algorithm~\cite{7445130} is proposed to decompose the non-convex data rate maximization problem into multiple tractable problems.
Alternatively,
to make the hybrid beamforming problem more tractable, an efficient approach is to transfer the maximization objective of the data rate to the minimization of the Euclidean distance between the fully-digital beamforming matrix and the hybrid beamforming matrix~\cite{1}. Through this transformation, the hybrid beamforming design problem can be simplified substantially.
An orthogonal matching pursuit (OMP) algorithm is attractive to solve the Euclidean distance minimization problem. 
To further reduce the complexity, various alternating optimization algorithms are proposed for the FC and AoSA architectures~\cite{1,7397861}, which solve the analog and digital beamforming matrices alternatively.

\subsection{Limitations and Remaining Problems}
As discussed before, the number of phase shifters in FC architecture equals to the product of the number of antennas and RF chains. By contrast, the AoSA architecture owns a limited number of phase shifters, i.e., equals to the number of antennas. Due to the exhaustive connection between the RF chains and antennas, the FC architecture can support a similar data rate to that of the optimal fully-digital beamforming architecture~\cite{1}. On the contrary, the data rate of the AoSA architecture is substantially compromised compared to the FC architecture, due to the partial connection between antennas and RF chains.

The FC and AoSA architectures can be considered as two extreme architectures for hybrid beamforming. 
The FC architecture owns a high data rate, while the power consumption and hardware complexity are high~\cite{111,8733134}. By contrast, the AoSA architecture owns low power consumption and hardware complexity, while sacrificing the data rate performance~\cite{7445130}. 
As studied in Sec.~\ref{section_challenges}-B, a balance of power consumption and data rate is critical for hybrid beamforming architecture in the THz band, which can not be achieved in the FC and AoSA architectures. 
Additionally, the low SDoF limitation has not been addressed in the FC and AoSA architectures. Employing these two architectures, the spatial multiplexing gain is still upper-bounded by the low SDoF, which further limits the data rate.
Moreover, these two architectures are based on phase shifter, where the beam squint problem remains as a critical issue to be addressed, as it could reduce received power and communication distance.
These limitations and remaining problems require careful design of novel THz-specific hybrid beamforming architectures, in order to bridge the gap between the theory and practical systems.

\section{THz-specific Hybrid Beamforming Architectures}
\label{section_THz_HBF}
In this section, we first introduce a novel widely-spaced multi-subarray (WSMS) hybrid beamforming architecture which can overcome the low SDoF limitation in the THz UM-MIMO systems. Second, we analyze a dynamic array-of-subarrays (DAoSA) hybrid beamforming architecture to balance the power consumption and data rate. 
Third, we study a true-time-delay-based (TTD-based) hybrid beamforming architecture, which has the potential to overcome the beam squint effect for THz communications with ultra-broad bandwidth. 
\subsection{THz WSMS Hybrid Beamforming Architecture}
\label{section_THz_HBF_WSMS}
\begin{figure}
	\centering
	\captionsetup{font={footnotesize}}
	\includegraphics[scale=0.39]{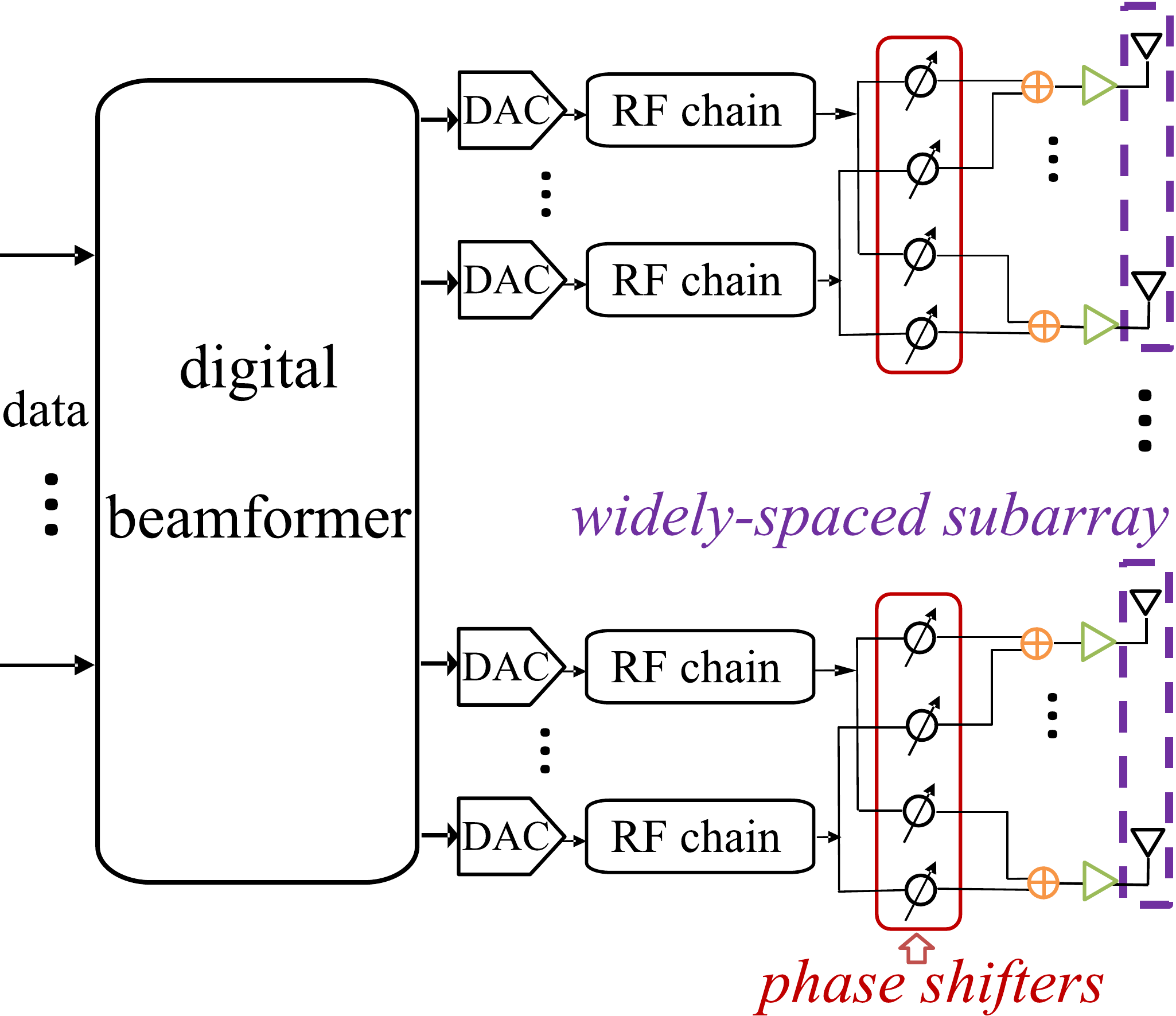}
	\caption{The THz WSMS hybrid beamforming architecture.}
	\label{Fig_architecture_WSMS}
	\vspace{-5.5mm}
\end{figure}
In THz UM-MIMO systems, multiple antennas can be used to explore the multiplexing gain that further enhances the data rate, in addition to beamforming.
Most of the studies on the FC and AoSA architectures assume that the antennas are separated by half of the wavelength, i.e., the antennas are critically-spaced~\cite{1,7397861,7445130,7389996}. With this critically-spaced antenna array, the spatial multiplexing gain is obtained from multipath of the channel, which is referred as \textit{inter-path multiplexing}~\cite{8356240}.
As analyzed in Sec. \ref{section_challenges}-A, due to the channel sparsity, the number of multipath in the THz channel is restricted, which results in a limited inter-path multiplexing gain and suppresses the data rate.

To further enhance the multiplexing gain, the WSMS hybrid beamforming architecture is promising by exploiting \textit{intra-path multiplexing} for THz UM-MIMO systems, as depicted in Fig.~\ref{Fig_architecture_WSMS}~\cite{8356240}.
The antennas array are composed of multiple subarrays. In each subarray, the antenna spacing is half of the wavelength, which is critically-spaced as in the existing FC and AoSA architectures. By contrast, the subarrays are separated over hundreds of wavelength, i.e., widely-spaced, which reduces the correlation between the subarrays. 
Interestingly, it has been analyzed that by setting the subarrays widely-spaced, the spherical-wave propagation needs to be considered for UM-MIMO channel modeling~\cite{8356240}.
As a result, one propagation path between the transmitted and received arrays can be decomposed into multiple sub-paths at the subarray level with distinct phases, which enhance the SDoF and lead to the additional intra-path multiplexing gain. 
Compared to the FC and AoSA architectures, the total multiplexing gain of the WSMS architecture is improved by a factor of the number of subarrays, by jointly utilizing the inter-path and intra-path multiplexing. Consequently, the data rate of the THz WSMS hybrid beamforming architecture is significantly improved compared to the FC and AoSA architectures.

\textit{\textbf{Algorithm design}:}
In the WSMS architecture,
since the subarrays are widely-spaced, the RF chain connected with one subarray can not connect to other subarrays, which results in a block-diagonal analog beamforming matrix. This block-diagonal-structured constraint brings difficulties to solve the analog and digital beamforming matrices. One method is leveraging the matrix decomposition idea to transfer this block-diagonal-structured problem into multiple FC hybrid beamforming problems, which can be solved by the existing FC hybrid beamforming algorithms.

\textit{\textbf{Remarks}:}
In the THz WSMS architecture, the widely-spaced subarrays are useful to overcome the low SDoF challenge in the THz sparse channel. The overall multiplexing gain is improved by a factor of the number of subarrays, which further improves the data rate significantly. However, due to the trade-off between the multiplexing and beamforming, the growth of multiplexing gain reduces the array gain, which influences the SNR as well as the data rate. Therefore, the number of widely-spaced subarrays needs to be carefully designed to strike a balance between the multiplexing and array gains.

\subsection{THz DAoSA Hybrid Beamforming Architecture}
\label{section_THz_HBF_DAoSA}
\begin{figure}
	\centering
	\captionsetup{font={footnotesize}}
	\includegraphics[scale=0.41]{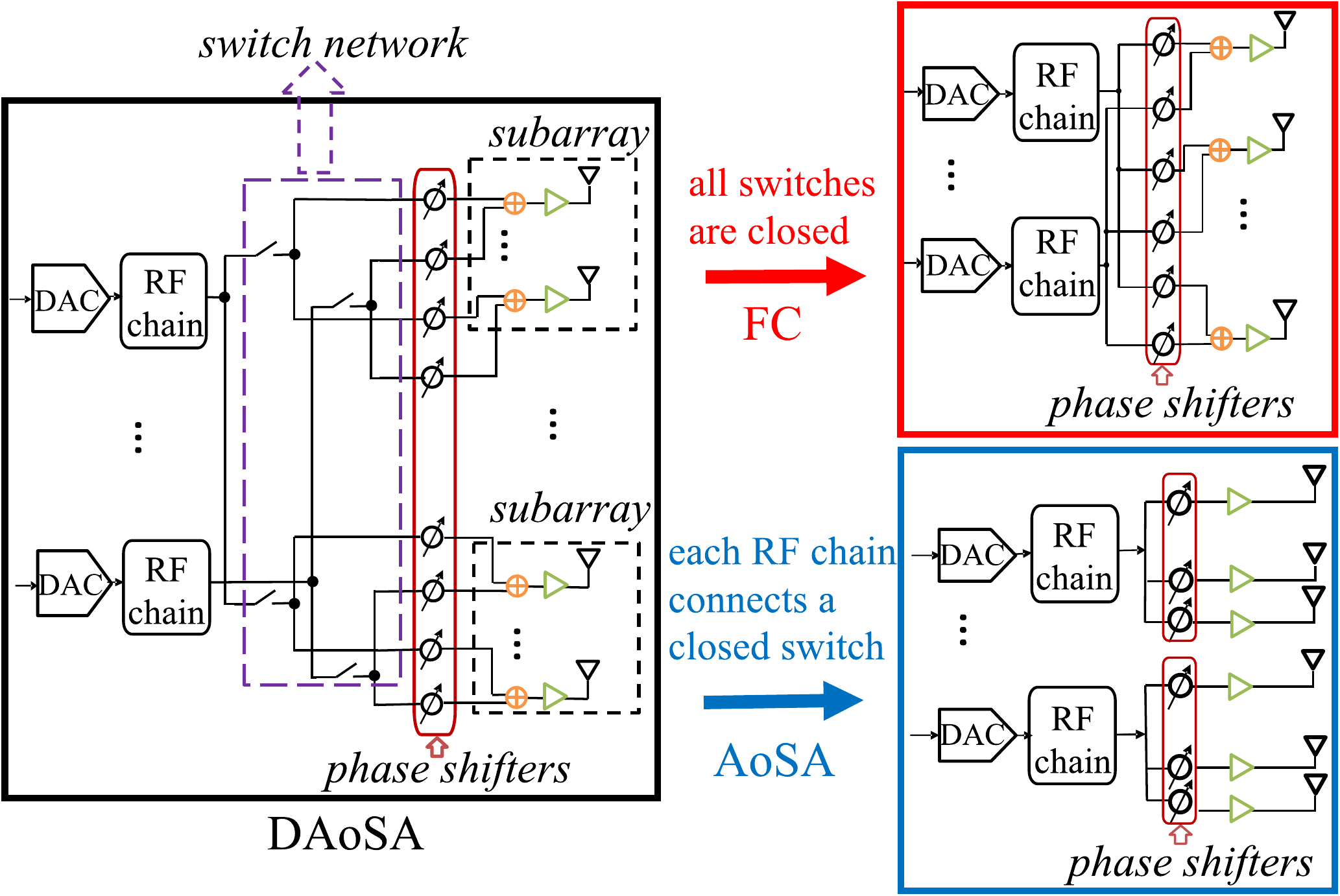}
	\caption{The analog part of the THz DAoSA hybrid beamforming architecture.}
	\label{Figure_architectures_DAoSA}
	\vspace{-2.5mm}
\end{figure}

We need to balance the power consumption and data rate of the THz hybrid beamforming architecture, inspired by the challenge of large-scale antenna array in THz UM-MIMO systems in Sec. \ref{section_challenges}-B. 
To this end, novel hybrid beamforming architectures with flexible hardware connection are proposed, e.g., the overlapped architecture~\cite{8733134} and the DAoSA architecture~\cite{111}.
Specifically, the architecture of DAoSA hybrid beamforming is presented in Fig.~\ref{Figure_architectures_DAoSA}. The antennas are divided into multiple subarrays. Switches are inserted between each RF chain and each subarray. Through controlling the state of the switch, i.e., open and closed, the connection between the RF chains and the subarrays can be intelligently adjusted. Particularly, the FC architecture is a special case of DAoSA with all switches closed. Conversely, the AoSA architecture is another special case of DAoSA, when each RF chain connects to one closed switch.
In the DAoSA design, the phase shifters that are connected with open switch are non-active and do not consume power.
Therefore, by dynamically designing switch connections, different levels of data rate and power consumption can be achieved, i.e., the power consumption and data rate can be balanced.

\textit{\textbf{Algorithm design}:}
The algorithm design in the DAoSA architecture includes two parts, namely, the switch network and beamforming matrices.
The switch network can be intelligently designed to maximize the energy efficiency or to minimize the consumed power while meeting the required data rate. The beamforming matrices are usually designed to maximize the data rate.
An alternating-selection method can alternatively calculate the switch network and beamforming matrices~\cite{111}. For the switch network part, an original intractable integer problem is transferred to a tractable sequence problem. 
Considering the beamforming matrices, due to the dynamical connection of the DAoSA architecture, the open switches force some entries in the analog beamforming matrix as zero. Since the state of the switches are adaptive, the positions of the zero entries in the analog beamforming matrix are usually irregular, which makes solving the analog and digital beamforming matrices intractable. For this sake, an element-by-element (EBE) algorithm can be used to design the analog and digital beamforming matrices~\cite{111}. Particularly, the EBE algorithm calculates the analog beamforming matrix element-by-element to address the irregular-structure constraint. 
\begin{table}
	\centering
	\captionsetup{font={footnotesize}}
	\caption{The wireless backhaul scenario is considered, in which a LoS path and a ground-reflection path compose the multipath channel that can be generated by the ray-tracing method~\cite{6998944}.}
	\begin{tabular}{|c|c|}
		\hline Distance between Tx and Rx&100m\\
		\hline Height of Tx and Rx&30m\\
		\hline Number of antennas at TX and Rx&1024\\
		\hline Central frequency&0.3 THz\\
		\hline Number of multipath& 2\\
		\hline
	\end{tabular}
	\label{simulation_parameters}
	\vspace{-4.5mm}
\end{table}

\textit{\textbf{Remarks}:}
Through intelligently determining the connection between subarrays and RF chains, the DAoSA architecture can balance the power consumption and data rate well. However, there are still some open problems in the DAoSA architecture. First, the algorithm design suffers from high complexity, since it contains both switch network design and beamforming design. One solution is using random matrix theory to determine the state of the switch network matrix with low complexity. Another potential direction is developing a unified framework to solve these two problems jointly to further reduce the complexity. Second, the hardware complexity of DAoSA architecture needs to be reduced. As presented in Fig.~\ref{Figure_architectures_DAoSA}, compared to the FC architecture, the DAoSA architecture has many additional switches, which could put a burden on practical implementation. 

\subsection{THz TTD-based Hybrid Beamforming}
\label{section_THz_HBF_TTD}
To solve the severe beam squint problem in the THz band, TTD-based hybrid beamforming is promising, where TTD is employed to substitute the phase shifter.
As analyzed in Sec.~\ref{section_challenges}-B, the beam squint problem arises from the fact that the required phase values in hybrid beamforming are related to the frequency, which is not achievable by the frequency-independent phase shifter.
This problem is particularly severe in wideband THz communications.
Fortunately, the TTD is frequency-dependent in its working band, i.e., the phase shift adjusted by TTD is proportional to the carrier frequency and can perfectly match the required phase shift of the wideband THz hybrid beamforming architectures~\cite{7959180}. 
Since the TTD-based hybrid beamforming is indeed a universal solution,
using TTD to adjust the phase can mitigate the beam squint problem in all FC, AoSA, WSMS, and DAoSA architectures.

\textit{\textbf{Algorithm design}:}
A codebook-based algorithm to design the TTD-based hybrid beamforming matrix is introduced in~\cite{7959180}. Based on the existing codebook-based algorithms for phase shifter, the proposed algorithm in~\cite{7959180} considers the frequency-dependent property of the TTD and redesigns the codebook. In general, the performance of the codebook-based algorithm relies on the size of the codebook. To achieve superior performance, the codebook size is usually large, which results in a high complexity.

\textit{\textbf{Remarks}:}
By using infinite-resolution TTD to substitute the phase shifter, the beam squint problem can be addressed perfectly without incurring array gain loss. However, the ideal infinite-resolution TTD or high-resolution TTD are fairly power-hungry and  suffer a high hardware complexity~\cite{7959180}. On the contrary, the energy-efficient low-resolution TTD is more suitable for practical systems. Note that the limitation of the resolution could cause a residual beam squint effect. The relationship between the resolution of TTD and the array gain loss needs to be analyzed.
Furthermore, low-complexity non-codebook-based algorithms are still lacked for THz TTD-based hybrid beamforming.

\section{Performance Evaluation}
\begin{figure}
	\centering
	\captionsetup{font={footnotesize}}
	\includegraphics[scale=0.41]{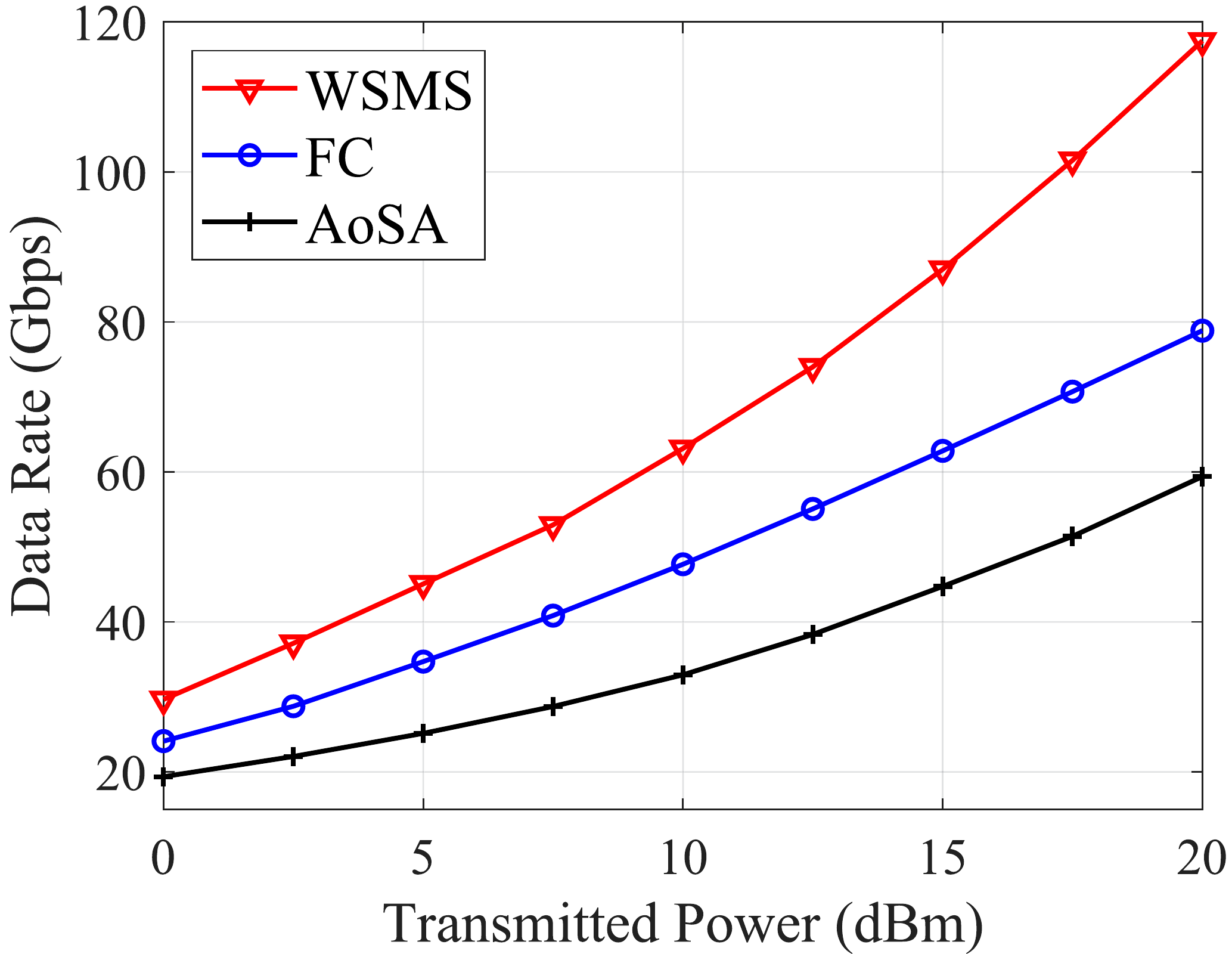}
	\caption{The data rate of the THz WSMS, FC and AoSA architectures. The bandwidth is 5 GHz and the number of RF chains equals to 8. Two widely-spaced subarrays are arranged in the WSMS architecture.}
	\label{Figure_SE_WSMS_FC_AoSA}
	\vspace{-2.5mm}
\end{figure}

We analyze the performance of the THz WSMS, DAoSA, TTD-based hybrid beamforming architectures in this section. The simulation parameters are given in TABLE~\ref{simulation_parameters}.

As presented in Fig.~\ref{Figure_SE_WSMS_FC_AoSA}, by equipping 2 widely-spaced subarrays, the SDoF and multiplexing gain in the WSMS architecture are doubled, which results in an enhanced data rate compared to the FC and AoSA architectures.
With 20 dBm transmitted power, the data rate of the THz WSMS hybrid beamforming architecture is 40~Gbps and 60~Gbps higher than those of the FC and AoSA counterparts, respectively, by addressing the low SDoF problem.

\begin{figure}
	\centering
	\captionsetup{font={footnotesize}}
	\includegraphics[scale=0.41]{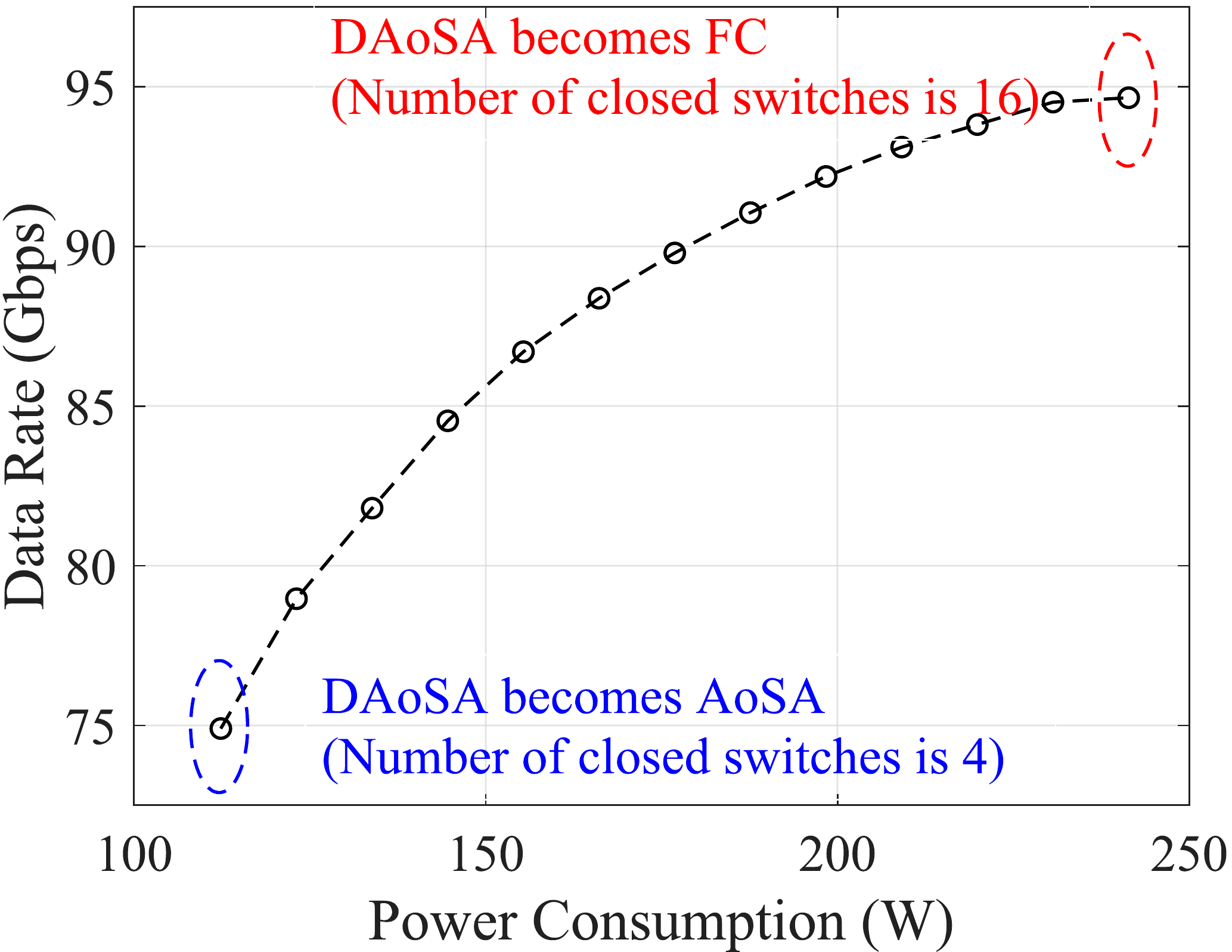}
	\caption{The data rate and power consumption of the THz DAoSA architecture. The bandwidth is 5 GHz and the transmit power equals to 20 dBm. The numbers of RF chains and switches are 4 and 16.}
	\label{Figure_SE_PC_DAoSA}
	\vspace{-2.5mm}
\end{figure}
Moreover, we analyze the data rate and power consumption of the DAoSA architecture in Fig.~\ref{Figure_SE_PC_DAoSA}. In the DAoSA architecture, with more closed switches, both the data rate and power consumption grow. Particularly, when all 16 switches are closed, the DAoSA becomes FC such that the data rate and power consumption reach the maximum values. In the other extreme case by closing 4 switches, the DAoSA is equivalent to AoSA, where the data rate reduces to the minimum value. Through intelligently controlling the switch network, i.e., the number of closed switches varies over $[4,16]$, the data rate and power consumption can be adjusted adaptively. Therefore, the THz DAoSA architecture can strike a good balance between the data rate and the power consumption.

\begin{figure}
	\centering
	\captionsetup{font={footnotesize}}
	\includegraphics[scale=0.41]{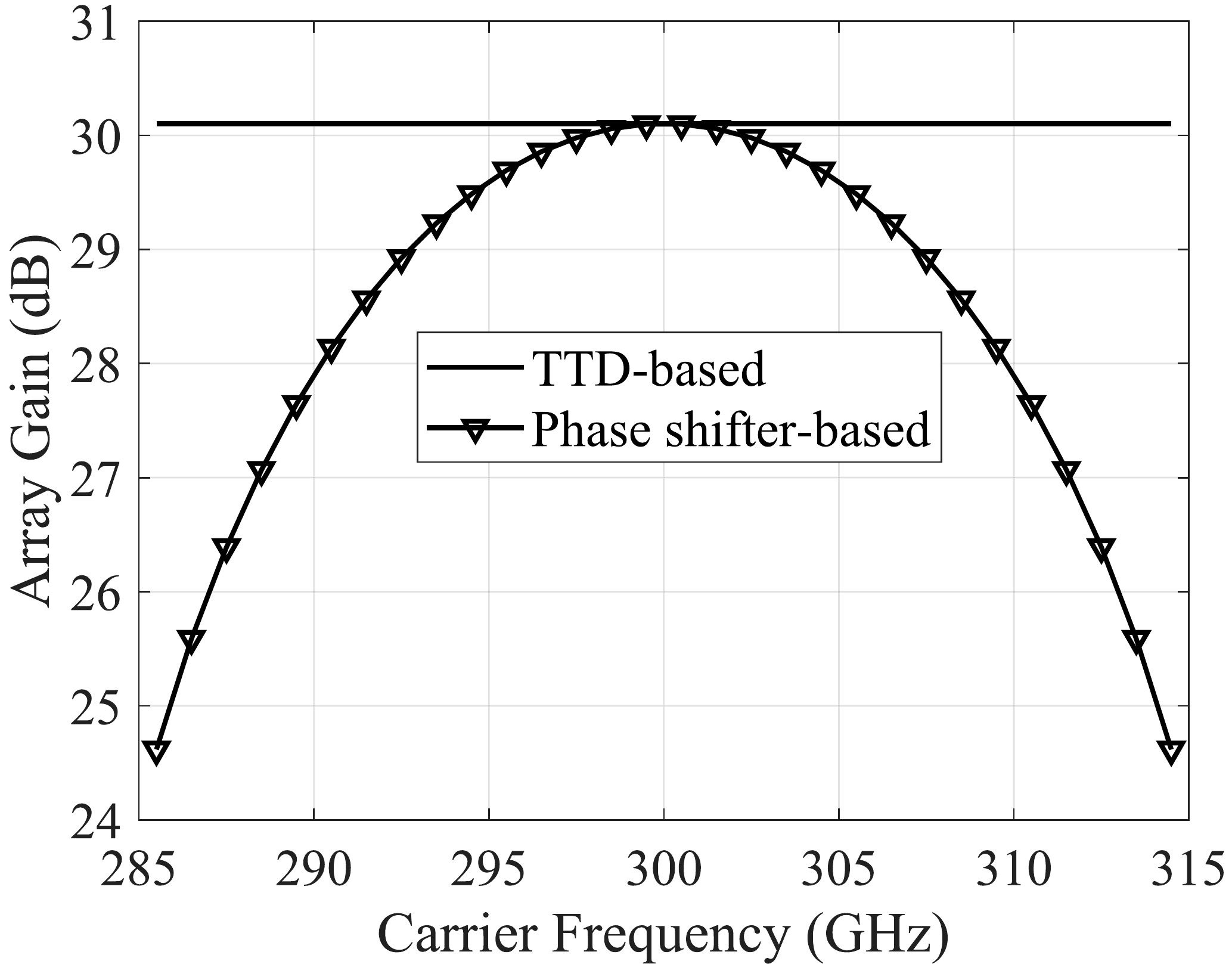}
	\caption{The array gain of the TTD-based and phase shifter-based FC architecture. The bandwidth is 30 GHz. The number of subcarriers is 30. The steering azimuth and elevation angles are $60^\circ$ and $10^\circ$.}
	\label{Figure_arraygain}
	\vspace{-2.5mm}
\end{figure}

Next, we compare the array gain of the FC architecture based on TTD and phase shifter, as illustrated in Fig.~\ref{Figure_arraygain}. For different carrier frequencies, the FC architecture that employs TTD to adjust the phase can attain a constant array gain.
By contrast, while using phase shifter, the targeted array gain is reached only at the central frequency. As the frequency deviates away from the central frequency, the array gain reduces with the deviation. 
In particular, the beam squint problem causes an array gain loss as large as 5.49~dB. 
Consequently, the use of TTD is beneficial for wideband THz hybrid beamforming design, to mitigate the beam squint problem and guarantee a constant high array gain.

\section{Open Problems and Potential Research Directions}
\label{section_problem}

\subsection{Hardware-efficient Design}
As the frequency grows to the THz band, the hardware challenges increase drastically. For instance, the high-resolution phase shifter, TTD, and DAC/ADC are still difficult to produce. Consequently, the THz hybrid beamforming architectures and algorithms by accounting for low-resolution phase shifter, TTD, and DAC/ADC are urgently needed for practical implementation. 
Another potential research direction is to study lens array to realize the THz hybrid beamforming, in which an electromagnetic lens is adopted to focus the THz wave, and
a matching antenna array is placed on the focal surface of the lens. By employing the lens array, the phase shifters or TTDs are not needed, which reduces the hardware complexity substantially. However, extensive switches are required to select beams from ultra-massive antennas, which increases the hardware complexity. Therefore, investigation on hardware-efficient lens array architectures and algorithms are needed.

\subsection{Influence of Imperfect Channel State Information}
Since the number of antennas is extremely high, the THz UM-MIMO channel is a high-dimensional matrix, which increases difficulties to obtain perfect channel state information (CSI). Additionally, with ultra-sharp beam generated by THz UM-MIMO systems, the performance degradation brought by imperfect CSI and beam misalignment is significant. Therefore, on one hand, super-resolution channel and angle estimation methods for THz hybrid beamforming architectures need to be developed. One potential solution is using subspace-tracking method to enhance the resolution of the off-grid MUSIC algorithm. On the other hand, the investigation of THz hybrid beamforming algorithms that are robust to imperfect CSI is needed. The probabilistic approach is one attractive idea to develop the robust THz hybrid beamforming algorithm~\cite{Yuan2020TCOM}.

\subsection{Deep Learning Algorithms for Hybrid Beamforming}
Although extensive hybrid beamforming algorithms have been investigated, their computational complexity is usually on the order of the number of antennas or even the square of the number of antennas~\cite{1,7445130,111,7397861,7389996}. For THz hybrid beamforming with ultra-massive antennas, e.g., 1024 antennas, the computational complexity is considerably high.
Recently, deep learning (DL) methods have drawn much attention for solving high-complexity physical communication and resource allocation problems.
In light of this, the authors in~\cite{9112250} propose to use the deep reinforcement learning approach to solve the hybrid beamforming problem at mmWave frequencies.
When applying DL algorithms to THz hybrid beamforming, the unique features and challenges elaborated in Sec.~\ref{section_challenges} need to be carefully considered.

\subsection{Joint Active and Passive Beamforming}
As analyzed in Sec. \ref{section_challenges}-A, blockage is a severe constraint in THz communications.
When the LoS path is blocked, the remaining paths might be too weak to support high data rate and the connection may even fall into outage. 
To address the blockage concern, joint active and passive beamforming that employs both UM-MIMO at the transmitter and receiver as well as intelligent reflecting surface (IRS) in the propagation environment is an interesting direction to explore, particularly for THz communications~\cite{8936989}.
The hybrid beamforming architecture performs active beamforming at the transmitter and receiver to improve the strength of the signal and the data rate. 
Meanwhile, IRS composed by a large number of reconfigurable passive elements can provide passive beamforming, to adjust the phase of the incident signal and collaboratively
change the direction of the reflected signal. Through the passive beamforming performed by IRS, the strength of the reflected signal is improved substantially, which enhances the robustness of the THz communications when the LoS path is blocked. The performance analysis and joint design of the active and passive beamforming algorithm in the THz band are still lacked, which need further studies.

\section{Conclusion}
\label{section_conclusion}
In this paper, we analyze the challenges and features of the THz hybrid beamforming design, including the low SDoF limitation, the blockage issue, the large scale antenna array constraint, the beam squint effect, and the spherical-wave propagation problem. Then, we introduce two traditional FC and AoSA hybrid beamforming architectures and investigate their limitations when being applied to THz band. Furthermore, we analyze three THz-specific hybrid beamforming architectures, i.e., THz WSMS, DAoSA, and TTD-based architectures. Simulation results are provided to validate their data rate, power consumption, and array gain for THz communications. More importantly, multiple open problems and potential research directions are discussed for THz hybrid beamforming design.

\bibliographystyle{IEEEtran}
\bibliography{IEEEabrv,references}

\balance
\end{document}